\documentclass[sigconf]{acmart}

\usepackage[final]{showlabels}
\usepackage{booktabs}
\usepackage{amsfonts}
\usepackage{amsmath}
\usepackage{graphicx}
\usepackage{balance}
\usepackage[T1]{fontenc}
\usepackage{subfigure}
\usepackage{algorithm2e}
\usepackage{amsthm}
\usepackage{hyperref}
\usepackage{xcolor}
\usepackage{pifont}
\usepackage{multirow}
\usepackage{array}
\usepackage[inline]{enumitem}
\usepackage{bbding}
\usepackage{wasysym}

\usepackage[skip=0.5\baselineskip]{caption}

\hypersetup{
    colorlinks=false,
    linkcolor={red!20!black},
    citecolor={green!20!black},
    urlcolor={blue!20!black}
}

\copyrightyear{2019}
\acmYear{2019}
\setcopyright{iw3c2w3}
\acmConference[WWW '19]{Proceedings of the 2019 World Wide Web Conference}{May 13--17, 2019}{San Francisco, CA, USA}
\acmBooktitle{Proceedings of the 2019 World Wide Web Conference (WWW '19), May 13--17, 2019, San Francisco, CA, USA}
\acmPrice{}
\acmDOI{10.1145/3308558.3313657}
\acmISBN{978-1-4503-6674-8/19/05}

\newcommand{\spara}[1]{\smallskip\noindent{\bf #1}}

\newcommand{\rurl}[1]{\href{http://#1}{#1}}
\newcommand{\cmark}{\ding{51}}
\newcommand{\xmark}{\ding{55}}
\newcommand{\rot}[1]{\rlap{\rotatebox{40}{#1}~}}

\title[SciLens: Evaluating the Quality of Scientific News]{SciLens: Evaluating the Quality of Scientific News Articles \mbox{Using Social Media and Scientific Literature Indicators}}
\author{Panayiotis Smeros}
\affiliation{%
  \institution{\'Ecole Polytechnique F\'ed\'erale de Lausanne (EPFL)}
  \city{Lausanne}
  \state{Switzerland}
}
\email{panayiotis.smeros@epfl.ch}

\author{Carlos Castillo}
\affiliation{%
	\institution{Universitat Pompeu Fabra (UPF)}
	\city{Barcelona}
	\state{Catalunya, Spain}
}
\email{carlos.castillo@upf.edu}

\author{Karl Aberer}
\affiliation{%
	\institution{\'Ecole Polytechnique F\'ed\'erale de Lausanne (EPFL)}
	\city{Lausanne}
	\state{Switzerland}
}
\email{karl.aberer@epfl.ch}

\renewcommand{\shortauthors}{P. Smeros et al.}

\begin{document}

% !TEX root = main.tex

\begin{abstract}
This paper describes, develops, and validates SciLens, a method to evaluate the quality of scientific news articles.
The starting point for our work are structured methodologies that define a series of quality aspects for manually evaluating news.
Based on these aspects, we describe a series of indicators of news quality. According to our experiments, these indicators help non-experts evaluate more accurately the quality of a scientific news article, compared to non-experts that do not have access to these indicators.
Furthermore, SciLens can also be used to produce a completely automated quality score for an article, which agrees more with expert evaluators than manual evaluations done by non-experts.
One of the main elements of SciLens is the focus on both content and context of articles, where context is provided by (1) explicit and implicit references on the article to scientific literature, and (2) reactions in social media referencing the article.
We show that both contextual elements can be valuable sources of information for determining article quality.
The validation of SciLens, done through a combination of expert and non-expert annotation, demonstrates its effectiveness for both semi-automatic and automatic quality evaluation of scientific news.
\end{abstract}

%TODO: ACM categories

 %TODO: keywords

\maketitle
% !TEX root = main.tex

\section{Introduction}\label{sec:introduction}

Scientific literacy is broadly defined as a knowledge of basic scientific facts and methods.
Deficits in scientific literacy are endemic in many societies, which is why understanding, measuring, and furthering the public understanding of science is important to many scientists~\cite{bauer2007can}.

Mass media can be a potential ally in fighting scientific illiteracy. Reading scientific content has been shown to help align public knowledge of scientific topics with the scientific consensus, although in highly politicized topics it can also reinforce pre-existing biases~\cite{hart2015public}.
There are many ways in which mass media approaches science, and even within the journalistic practice there are several sub-genres.
Scientific news portals, for instance, include most of the categories of articles appearing traditionally in newspapers~\cite{foust2017} such as  \textit{editorial}, \textit{op-ed}, and (less frequently) \textit{letters to the editor}. % articles on which, editors, journalists and the public audience respectively, can express their opinion on a particular scientific subject.
The main category of articles, however, are scientific \textit{news} articles, where journalists describe scientific advances. %, ideally through an objective and transparent methodology.
%Since the first categories of articles are tightly coupled with the opinion of the author, they appear to be less structured and thus it is harder to assess their quality.
%This is the category in which we focus our attention.

%\spara{Scientific news articles vs ordinary news articles.}
%
Scientific news articles have many common characteristics with other classes of news articles; for instance, they follow the well-known \textit{inverted pyramid} %\footnote{\url{https://en.wikipedia.org/wiki/Inverted_pyramid_(journalism)}}
style, where the most relevant elements are presented at the beginning of the text.
However, they also differ in important ways.
Scientific news are often based on findings reported in scientific journals, books, and talks, which are highly specialized.
The task of the journalist is then to \textit{translate} these findings to make them understandable to a non-specialized, broad audience.
By necessity, this involves negotiating several trade-offs between desirable goals that sometimes enter into conflict, including appealing to the public and using accessible language, while at the same time accurately representing research findings, methods, and limitations~\cite{palmerini_2017}. % Also promoting science and educating
%
%\paragraph{Sources of scientific articles}
%According to the principles of journalism ... journalists have to be transparent regarding their sources.
%For scientific journalists, this task is not very trivial.

The resulting portrayal of science in news varies widely in quality.
For example, the website ``Kill or Cure?''\footnote{\url{http://kill-or-cure.herokuapp.com}} has reviewed over 1,200 news stories published by The Daily Mail (a UK-based tabloid) finding headlines pointing to 140 substances or factors that cause cancer (including obesity, but also Worcestershire sauce), 113 that prevent it (including garlic and green tea), and 56 that both cause and prevent cancer (including rice).
Evidently, news coverage of cancer research that merely seeks to classify every inanimate object into something that either causes or prevents cancer does not help to communicate effectively scientific knowledge on this subject.

\begin{figure}[t]
	\centering
	\includegraphics[width=\columnwidth]{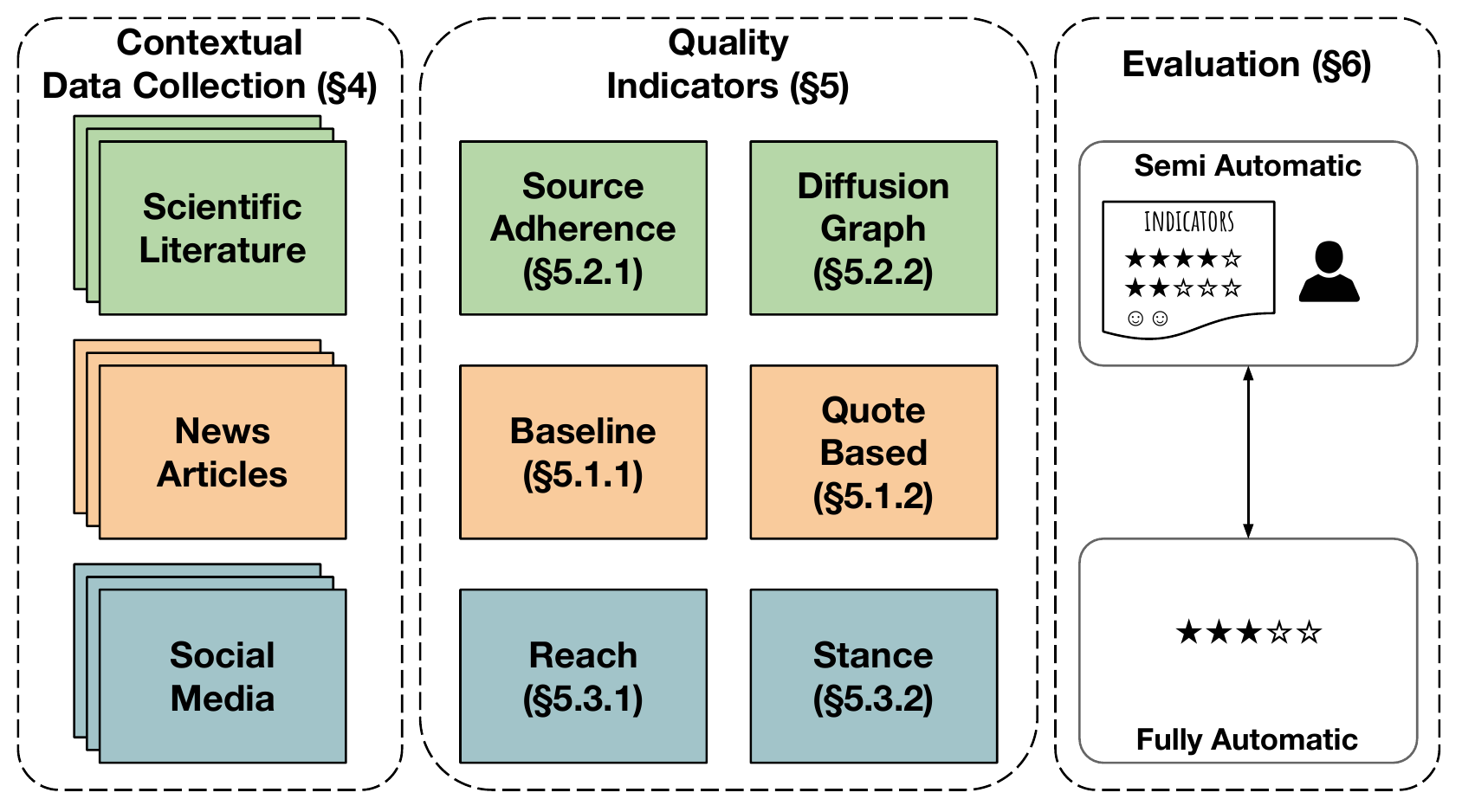}
	\caption{Overview of SciLens, including contextual data collection, quality indicators, and evaluation.}
	\label{fig:scilens_architecture}
\end{figure}

\spara{Our contribution.}
In this paper we describe \emph{SciLens,} a method for evaluating the quality of scientific news articles.
The technical contributions we describe are the following:
\begin{itemize}[nosep,leftmargin=1em,labelwidth=*,align=left]
\item a framework, depicted in Figure~\ref{fig:scilens_architecture}, for semi-automatic and automatic article quality evaluation (\S\ref{sec:scilens});
\item a method for contextual data collection that captures the contents of an article, its relationship with the scientific literature, and the reactions it generates in social media (\S\ref{sec:corpus});
\item a series of automatically-computed \emph{quality indicators} describing:
    \begin{itemize}
        \item the content of a news article, where we introduce a method to use quotes appearing on it as quality indicators (\S\ref{subsec:methods-content}),
        \item the relationship of a news article with the scientific literature, where we introduce content-based and graph-based similarity methods (\S\ref{subsec:methods-papers}), and
        \item the social media reactions to the article, where we introduce a method to interpret their stance (supporting, commenting, contradicting, or questioning) as quality signals  (\S\ref{subsec:methods-social});
    \end{itemize}
\item an experimental evaluation of our methods involving experts and non-experts (\S\ref{sec:results}).
\end{itemize}

%\smallskip
%\noindent The next section summarizes related work; Section~\ref{sec:conclusions} presents conclusions, limitations, and outlines future work.

% !TEX root = main.tex

\begin{table*}[t]
			\centering
			\caption{Summary of selected references describing techniques for evaluating news article quality.}
			\label{table:qa_techniques}
			\begin{tabular}{l*{12}{l}c}
				 &
				 \rot{Fact-checking portals} &
				 \rot{\citet{DBLP:conf/www/ShaoCFM16}} &
				 \rot{\citet{DBLP:conf/mir/BoididouPAK17}} &
				 \rot{\citet{DBLP:conf/www/PopatMSW17}} &
				 \rot{\citet{DBLP:conf/www/TambuscioRFM15}} &
				 \rot{\citet{doi:10.1371/journal.pone.0128193}} &
	 			 \rot{\citet{doi:10.1080/1461670X.2013.856670}}  &
	 			 \rot{\citet{DBLP:conf/www/ZhangRMASGAVLRB18}} &
	 			 \rot{\citet{DBLP:journals/sigkdd/ShuSWTL17}} &
	 			 \rot{\citet{DBLP:conf/www/Kumar0L16}} &
				 \rot{\citet{doi:10.2196/jmir.7579}} &
				 \rot{\citet{doi:10.1371/journal.pone.0127848}} &
				 \textbf{SciLens}\\
				 \hline
				Automatic assessment & \xmark & \cmark  & \cmark & \cmark & \cmark & \cmark & \xmark & \xmark & \cmark & \cmark & \xmark  & \xmark  & \cmark \\
				\hline
				No ground-truth needed & \cmark & \xmark & \xmark & \xmark & \xmark & \xmark & \cmark & \cmark & \xmark & \cmark & \cmark & \cmark & \cmark \\
				\hline
				Uses article content  & \cmark & \xmark  & \xmark & \cmark & \xmark & \cmark & \cmark & \cmark & \cmark & \cmark & \cmark & \cmark & \cmark \\
				\hline
				Uses reactions from social media  & \xmark & \cmark  & \cmark & \cmark & \cmark & \xmark & \xmark & \cmark & \cmark & \xmark & \xmark & \xmark & \cmark \\
				\hline
				Uses referenced scientific literature & \cmark & \xmark  & \xmark & \xmark & \xmark & \xmark & \xmark & \cmark & \xmark & \xmark & \cmark & \cmark & \cmark \\
				\hline
				Domain-agnostic & \cmark & \cmark  & \cmark & \cmark & \cmark & \xmark & \cmark & \cmark & \cmark & \cmark & \xmark & \xmark & \cmark \\
				\hline
				Web-scale & \xmark & \cmark & \cmark & \cmark & \cmark & \cmark & \xmark & \xmark & \cmark & \cmark & \xmark & \xmark & \cmark \\
				\hline

			\end{tabular}
\end{table*}

\section{Related Work}\label{sec:RelatedWork}

In this section, we present background information that frames our research (\S\ref{subsec:background-scientific-news}), previous work on evaluating news quality (\S\ref{subsec:RelatedWork-quality}), and methods to extract quality indicators from news articles (\S\ref{subsec:relwork-indicator}).
This is a broad research area where results are scattered through multiple disciplines and venues; our coverage is by no means complete.

\subsection{Background on Scientific News}\label{subsec:background-scientific-news}

A starting point for understanding communication of science has historically been the ``deficit model,'' in which the public is assumed to have a deficit in scientific information that is addressed by science communication~(see, e.g., \citet{doi:10.1088/0963-6625/3/1/001}).
In a simplified manner, scientific journalism, as practiced by professional journalists as well as science communicators and bloggers from various backgrounds, can be seen as a translation from a discourse inside scientific institutions to a discourse outside them.
However, there are many nuances that make this process much more than a simple translation. For instance, \citet{myers2003discourse}, among others, notes that
\begin{enumerate*}[label=(\roman*)]
\item in many cases the gulf between experts and the public is not as large as it may seem, as many people may have some information on the topic;
\item there is a continuum of popularization through different genres, i.e., science popularization is a matter of degree;
and
\item the scientific discourse is intertwined with other discourses, including the discussion of political and economic issues.
\end{enumerate*}

Producing a high-quality article presenting scientific findings to the general public is unquestionably a challenging task, and often there is much to criticize about the outcome.
In the process of writing an article, ``information not only changes textual form, but is simplified, distorted, hyped up, and dumbed down''~\cite{myers2003discourse}.
Misrepresentation of scientific knowledge by journalists has been attributed to several factors, including ``a tendency to sensationalize news, a lack of analysis and perspective when handling scientific issues, excessive reliance on certain professional journals for the selection of news, lack of criticism of powerful sources, and lack of criteria for evaluating information''~\cite{semir2000scientific}.

In many cases, these issues can be traced to journalists adhering to journalistic rather than scientific norms.
According to~\citet{dunwoody2014science}, this includes
\begin{enumerate*}[label=(\roman*)]
\item a tendency to favor conflict, novelty, and similar news values;
\item a compromise of accuracy by lack of details that might be relevant to scientists, but that journalists consider uninteresting and/or hard to understand for the public;
and
\item a pursuit of ``balance'' that mistakenly gives similar coverage to consensus viewpoints and fringe theories.
\end{enumerate*}
Journalists tend to focus on events or episodic developments rather than long-term processes, which results in preferential coverage to initial findings even if they are later contradicted, and little coverage if results are disconfirmed or shown to be wrong~\cite{doi:10.1371/journal.pone.0172650}.
Furthermore, news articles typically do not include caveat/hedging/tentative language, i.e., they tend to report scientific findings using a language expressing certainty, which may actually have the opposite effect from what is sought, as tentative language makes scientific reporting more credible to readers~\cite{jensen2008scientific}.

\subsection{Evaluation of Quality of News}\label{subsec:RelatedWork-quality}

There are many approaches for evaluating the quality of articles on the web; we summarize some of these approaches in Table~\ref{table:qa_techniques}.

\spara{Manual Evaluation.}
The simplest approach for evaluating news article quality relies on the manual work of domain experts.
This is a highly subjective task, given that quality aspects such as credibility are to a large extent perceived qualities, made of many dimensions~\cite{DBLP:conf/chi/FoggT99}.
In the health domain, evaluations of news article quality have been undertaken for both general health topics~\cite{doi:10.2196/jmir.7579} and specific health topics such as Pancreatic Cancer~\cite{doi:10.1371/journal.pone.0127848}.

Independent, non-partisan \emph{fact-checking portals} perform manual content verification at large scale, typically employing a mixture of professional and volunteer staff. They can cover news articles on general topics (e.g., \rurl{Snopes.com}) or specific topics such as politics (e.g., \rurl{PolitiFact.com}).
In the case of science news, \rurl{ClimateFeedback.org} is maintained by a team of experts on climate change with the explicit goal of helping non-expert readers evaluate the quality of news articles reporting on climate change. Each evaluated article is accompanied by a brief review and an overall quality score.
Reviews and credibility scores from fact-checking portals have been recently integrated with search results~\cite{google2017} and social media posts~\cite{fb2017} to help people find accurate information.
Furthermore, they are frequently used as a ground truth to build systems for rumor tracking~\cite{DBLP:conf/www/ShaoCFM16}, claim assessment~\cite{DBLP:conf/www/PopatMSW17}, and fake multimedia detection~\cite{DBLP:conf/mir/BoididouPAK17}.
Articles considered by fact-checking portals as misinformation have been used as ``seeds'' for diffusion-based methods studying the spread of misinformation~\cite{DBLP:conf/www/TambuscioRFM15}.

Our approach differs from previous work because it is completely automated and does not need to be initialized with labels from expert- or crowd-curated knowledge bases. % bases, not based
For instance, in the \textit{diffusion graph}, which is the graph we construct during contextual data collection (\S\ref{sec:corpus}) from social media posts and scientific papers,  we do not need prior knowledge on the quality of different nodes.

\spara{Automatic and Semi-Automatic Evaluation.}
Recent work has demonstrated methods to automate the extraction of signals or indicators of article quality.
These indicators are either expressed at a conceptual level~\cite{doi:10.1080/1461670X.2013.856670} (e.g, \textit{balance of view points}, \textit{respect of personal rights}) or operationalized as features that can be computed from an article~\cite{DBLP:conf/www/ZhangRMASGAVLRB18} (e.g., \textit{expert quotes} or \textit{citations}).
\citet{DBLP:journals/sigkdd/ShuSWTL17} describe an approach for detecting fake news on social media based on social and content indicators.
\citet{DBLP:conf/www/Kumar0L16} describe a framework for finding hoax Wikipedia pages mainly based on the author's behavior and social circle, while \citet{doi:10.1371/journal.pone.0128193} use Wikipedia as ground truth for testing the validity of dubious claims.
\citet{DBLP:conf/emnlp/BalyKAGN18} describe site-level indicators that evaluate an entire website instead of individual pages.

Our work differs from these by being, to the best of our knowledge, the first work that analyzes the quality of a news article on the web combining its own content with context that includes social media reactions and referenced scientific literature.
We provide a framework, \emph{SciLens}, that is scalable and generally applicable to any technical/scientific context at any granularity (from a broad topic such as ``health and nutrition'' to more specific topics such as ``gene editing techniques'').

\subsection{Indicator Extraction Techniques}
\label{subsec:relwork-indicator}

Our method relies on a series of indicators that can be computed automatically, and intersects previous literature that describes related indicators used to evaluate article quality or for other purposes.

\spara{Quote Extraction and Attribution.}\label{subsec:relwork-quote-extraction} % Will be referencing the parent subsection
The most basic approach to quote extraction is to consider that a quote is a ``block of text within a paragraph falling between quotation marks''~\cite{DBLP:conf/aaai/ElsonM10, pouliquen2007automatic}.
Simple regular expressions for detecting quotes can be constructed~\cite{DBLP:conf/emnlp/OKeefePCKH12, DBLP:conf/nodalida/SalwayMHR17}.
\citet{DBLP:conf/icwsm/PavlloP018} leverages the redundancy of popular quotes in large news corpora (e.g., highly controversial statements from politicians that are intensely discussed in the press) for building unsupervised bootstrapping models, while \citet{DBLP:conf/emnlp/ParetiOKCK13} and \citet{DBLP:conf/eacl/JurafskyCMF17} train supervised machine learning models using corpora of political and literary quotes
(Wikiquote, \url{https://www.wikiquote.org}, is such a corpus that contains general quotes).

Our work does not rely on simple regular expressions, such as syntactic patterns combined with quotations marks, which in our preliminary experiments performed poorly in quote extraction from science news; instead we use regular expressions based on classes of words.
We also do not use a supervised approach as there is currently no annotated corpus for scientific quote extraction.
For the research reported on this paper, we built an information extraction model specifically for scientific quotes from scratch, i.e., a ``bootstrapping'' model, which is based on word embeddings. This is a commonly used technique for information extraction when there is no training data and we can manually define a few high-precision extraction patterns \cite{DBLP:conf/kdd/JinHS09}.

\spara{Semantic Text Similarity.}
One of the indicators of quality that we use is the extent to which the content of a news article represents the scientific paper(s) it is reporting about.
The Semantic Text Similarity task in Natural Language Processing (NLP) determines the extent to which two pieces of text are semantically equivalent.
This is a popular task in the \textit{International Workshop on Semantic Evaluation} (SemEval).
Three approaches that are part of many proposed methods over the last few years include:
\begin{enumerate*}[label=(\roman*)]
	\item \textit{surface-level similarity} (e.g., similarity between sets or sequences of words or named entities in the two documents);
	\item \textit{context similarity} (e.g., similarity between document representations); and
	\item \textit{topical similarity}
\end{enumerate*}
\cite{DBLP:conf/semeval/HanMCT15, DBLP:conf/semeval/LiebeckPM016}.

In this paper, we adopt these three types of similarity, which we compute at the document, paragraph, and sentence level. The results we present suggest that combining different similarity metrics at different granularities results in notable improvements over using only one metric or only one granularity.

\spara{Social Media Stance Classification.}
Our analysis of social media postings to obtain quality indicators considers their \emph{stance}, i.e., the way in which posting authors position themselves with respect to the article they are posting about.
Stance can be binary (``for'' or ``against''), or be described by more fine-grained types (supporting, contradicting, questioning, or commenting)~\cite{DBLP:conf/ijcnlp/HasanN13}, which is what we employ in this work.
Stance classification of social media postings has been studied mostly in the context of online marketing~\cite{DBLP:conf/dawak/KonjengbamGKS18} and political discourse and rumors~\cite{DBLP:journals/ipm/ZubiagaKLPLBCA18}.

In our work, we build a new stance classifier based on textual and contextual features of social media postings and replies, annotated by crowdsourcing workers. To the best of our knowledge, there is no currently available corpus covering the scientific domain.
As part of our work, we release such corpus.

% !TEX root = main.tex

\section{SciLens Overview}
\label{sec:scilens}

The goal of SciLens is to help evaluate the quality of scientific news articles.
As Figure~\ref{fig:scilens_architecture} shows, we consider a contextual data collection, a computation of quality indicators, and an evaluation of the results.

\spara{Contextual Data Collection (\S\ref{sec:corpus}).}
First, we consider a set of keywords that are representative of a scientific/technical domain; for this paper, we have considered a number of key words and phrases related to health and nutrition.
Second, we extract from a social media platform (in this case, Twitter), all postings matching these keywords, as well as public replies to these postings.
Third, we follow all links from the postings to web pages, and download such pages; while the majority of them are news sites and blogs of various kinds, we do not restrict the collection by type of site at this point.
Fourth, we follow all links from the web pages to URLs in a series of pre-defined domain names from scientific repositories, academic portals and libraries, and universities.
Fifth, we clean-up the collection by applying a series of heuristics to de-duplicate articles and remove noisy entries.

\spara{Quality Indicators (\S\ref{sec:methods}).}
We compute a series of quality indicators from the content of articles, and from their referencing social media postings and referenced scientific literature.

Regarding the content of the articles, we begin by computing several content-based features described by previous work.
Next, we perform an analysis of quotes in articles, which are a part of journalistic practices in general and are quite prevalent in the case of scientific news. Given that attributed quotes are more telling of high quality than unattributed or ``weasel'' quotes, we also carefully seek to attribute each quote to a named entity which is often a scientist, but can also be an institution.

Regarding the scientific literature, we would like to know the strength of the connection of articles to scientific papers. For this, we consider two groups of indicators: content-based and graph-based.
Content-based indicators are built upon various metrics of text similarity between the content of an article and the content of scientific papers, considering that the technical vocabulary is unlikely to be preserved as-is in articles written for the general public.
Graph-based indicators are based on a diffusion graph in which scientific papers and web pages in academic portals are nodes connected by links. High-quality articles are expected to be connected through many short paths to academic sources in this graph.

Regarding social media postings, we measure two aspects: reach and stance. Reach is measured through various proxies for attention, that seek to quantify the impact that an article has in social media. The stance is the positioning of posting authors with respect to an article, which can be positive (supporting, or commenting on an article without expressing doubts), or negative (questioning an article, or directly contradicting what the article is saying).

\spara{Evaluation (\S\ref{sec:results}).}
We evaluate the extent to which the indicators computed in SciLens are useful for determining the quality of a scientific news article.
We consider that these indicators can be useful in two ways.
First, in a semi-automatic setting, we can show the indicators to end-users and ask them to evaluate the quality of a scientific news article; if users that see these indicators are better at this task that users that do not see them, we could claim that the indicators are useful.
Second, in a fully automatic setting, we can train a model based on all the indicators that we computed.
In both cases, the ground truth for evaluation is provided by experts in communication and science.

% !TEX root = main.tex

\section{Contextual Data Collection}
\label{sec:corpus}

The contextual data collection in our work seeks to capture all relevant content for evaluating news article quality, including referenced scientific papers and reactions in social media.
This methodology can be applied to any specialized or technical domain covered in the news, as long as:
\begin{enumerate*}[label=(\roman*)]
	\item media coverage in the domain involves ``translating'' from primary technical sources,
	\item such technical sources can be characterized by known host/domain names on the web, and
	\item social media reactions can be characterized by the presence of certain query terms.
\end{enumerate*}
Examples where this type of contextual data collection could be applied beyond scientific news include news coverage of specialized topics such as law or finance.

\begin{figure}[t]
	\centering
	\includegraphics[width=\columnwidth]{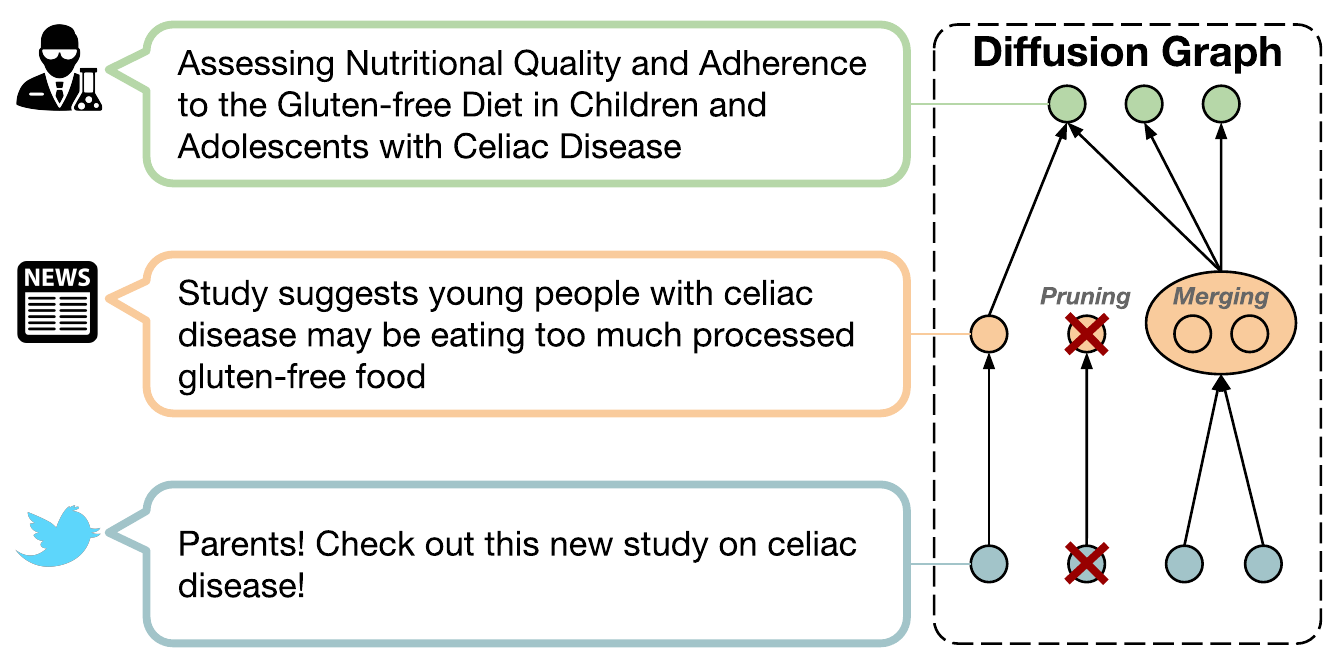}
	\caption{Contextual data collection, including social media postings, which reference a series of news articles, which cite one or more scientific papers. In our diffusion graph, paths that do not end up in a scientific paper or paths that contain unparsable nodes (e.g., malformed HTML pages) are \textit{pruned}, and articles with the same content in two different outlets (e.g., produced by the same news agency) are \textit{merged}.}
	\label{fig:scilens_architecture_crawling}
\end{figure}

We consider two phases:
a crawling phase, which starts from social media and then collects news articles and primary sources (\S\ref{subsec:corpus-crawling}), and
a pruning/merging phase, which starts from primary sources and prunes/de-duplicates articles and postings (\S\ref{subsec:corpus-pruning}).
This process is depicted in Figure~\ref{fig:scilens_architecture_crawling} and explained next.

\subsection{Crawling of Postings, Articles, and Papers}
\label{subsec:corpus-crawling}

The crawling phase starts with social media postings, which are identified as candidates for inclusion based on the presence of certain topic-related keywords in them.
In the case of this study, we selected ``health and nutrition'' as our main topic: this is among the most frequent topics in scientific news reporting, which is known to have a medical/health orientation~\cite{weitkamp2003british,badenschier2012issue,dunwoody2014science}.
The initial set of keywords was obtained from Nutrition Facts (\url{https://nutritionfacts.org/topics}), a non-commercial and non-profit website that provides scientific information on healthy eating.
The list contains over 2,800 keywords and key phrases such as ``HDL cholesterol,'' ``polyphenols'' and the names of hundreds of foods from ``algae'' to ``zucchini''.
We further expanded this list with popular synonyms from WordNet~\cite{DBLP:journals/cacm/Miller95}.

We harvest social media postings
from \rurl{DataStreamer.io} (formerly known as \rurl{Spinn3r.com}), covering a 5-year period from June 2013 through June 2018.
In this collection, we find $2.5M$ candidate postings matching at least one of our query terms from which we discard postings without URLs.
%
%We additionally collect all ``replies'' to them.

Next, we crawl the pages pointed to by each URL found in the remaining postings. These pages are hosted in a wide variety of domains, the majority being news outlets and blogging platforms.
We scan these pages for links to scientific papers, which we do identify by domain names.
We use a predefined list of the top-1000 universities as indicated by \rurl{CWUR.org} plus a manually curated list of open-access publishers and academic databases obtained from Wikipedia\footnote{\url{https://en.wikipedia.org/wiki/List_of_academic_databases_and_search_engines}} and expanded using the ``also visited websites'' functionality of \rurl{SimilarWeb.com}.
Overall, we obtained a diffusion graph of $\textbf{2.4M}$ nodes and $\textbf{3.7M}$ edges.

\subsection{Pruning and Merging}
\label{subsec:corpus-pruning}

The initial data collection described in \S\ref{subsec:corpus-crawling} is recall-oriented. Now, we make it more precise by pruning and merging items.

\spara{Pruning.}
During the pruning phase, we keep in our collection only documents that we managed to successfully download and parse (e.g., we discard malformed HTML pages and PDFs).
We also prune paths that do not end up in a scientific paper i.e., articles that do not have references and all the tweets that point to these articles.
This phase helps us eliminate most of the noisy nodes of the diffusion graph that were introduced due to the generic set of seed keywords that we used in the crawling phase (\S\ref{subsec:corpus-crawling}).

\spara{Merging.}
We notice a large number of duplicate articles across news outlets, which we identify by text similarity i.e, by cosine similarity of more than $90\%$ between the bag-of-words vectors representing the articles.
This happens when one outlet re-posts an article originally published in another outlet, or when both syndicate from the same news agency.
Once we find such duplicates or near-duplicates, we keep only one of them (the one having more out-links, breaking ties arbitrarily) and remove the redundant ones.
Social media postings that point to the duplicates are re-wired to connect to the one that survived after merging, hence we do not lose a potentially important signal of article quality.

\medskip
The resulting collection is large and mostly composed of elements that are closely related to the topic of health and nutrition: $\textbf{49K}$ social media postings, $\textbf{12K}$ articles (most of them in news sites and blogs), and $\textbf{24K}$ scientific links (most of them peer-reviewed or grey-literature papers).
Even after pruning, our collection is possibly more comprehensive than the ones used by systems used to appraise the impact of scientific papers.
For instance, when compared to \rurl{Altmetric.com} \cite{DBLP:journals/lp/AdieR13} we find that our collection has more links to scientific papers than what Altmetric counts. In their case, referencing articles seem to be restricted to a controlled list of mainstream news sources, while in our case we often find these references plus multiple references from less known news sources, blogs, and other websites.
%
%We consider the inclusion of the latter as an important step considering the popularity of less known websites, and their importance considering the current debate around ``fake news.''

% !TEX root = main.tex

\section{Quality Indicators}
\label{sec:methods}

We compute indicators from the content of news articles (\S\ref{subsec:methods-content}),
from the scientific literature referenced in these articles (\S\ref{subsec:methods-papers}), and
from the social media postings referencing them (\S\ref{subsec:methods-social}).
The full list of indicators is presented on Table~\ref{table:indicators_full}.

\subsection{News Article Indicators}
\label{subsec:methods-content}

These indicators are based on the textual content of a news article.

\subsubsection{Baseline Indicators}
\label{subsubsec:methods-content-baseline}

As a starting point, we adopt a large set of  content-based quality indicators described by previous work.
These indicators are:
\begin{enumerate*}[label=(\roman*)]
	\item title deceptiveness and sentiment: we consider if the title is ``clickbait'' that  oversells the contents of an article in order to pique interest \cite{DBLP:conf/ijcai/WeiW17, SaurabhMathur2017};
	\item article readability: indicator of the level of education someone would need to easily read and understand the article \cite{flesch1948new};
	and
	\item article length and presence of author byline \cite{DBLP:conf/www/ZhangRMASGAVLRB18}. % "signature" of an author in a newspaper is called the "byline"
\end{enumerate*}

\subsubsection{Quote-Based Indicators}
\label{subsubsec:methods-content-quotes}

Quotes are a common and important element of many scientific news articles. While selected by journalists, they provide an opportunity for experts to directly present their viewpoints in their own words~\cite{conrad1999uses}.
However, identifying quotes in general is challenging, as noted by previous work (\S\ref{subsec:relwork-quote-extraction}). In the specific case of our corpus, we observe that they are seldom contained in quotation marks in contrast to other kinds of quotes (e.g., political quotes \cite{pouliquen2007automatic}).
We also note that each expert quoted tends to be quoted once, which makes the problem of \emph{attributing} a quote challenging as well.

\begin{figure}[t]
	\centering
	\includegraphics[width=\columnwidth]{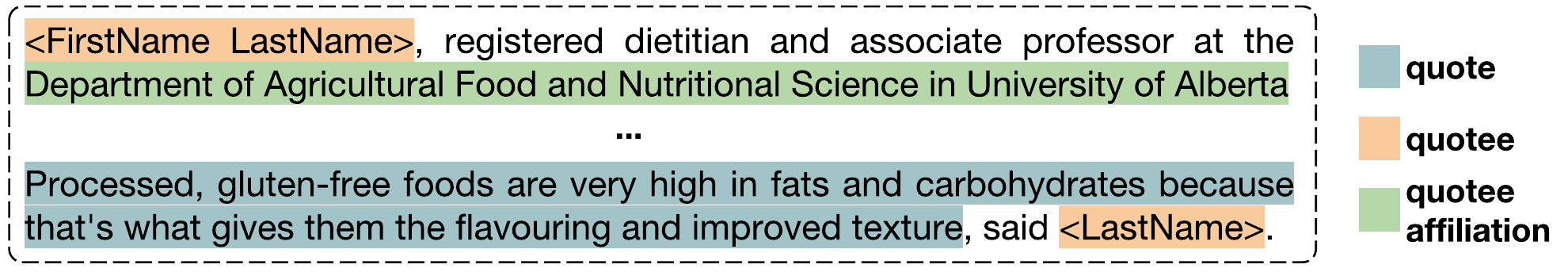}
	\caption{Example of quote extraction and attribution (best seen in color). Quotee has been anonymized.}
	\label{fig:quote_example}
\end{figure}

\spara{Quote Extraction Model.}
To extract quotes we start by addressing a classification problem at the level of a sentence: we want to distinguish between quote-containing and non-containing sentences.
To achieve this, we first select a random sample from our dataset, then manually identify quote patterns, and finally, we generalize automatically these patterns to cover the full dataset.
As we describe in the related work section (\S\ref{subsec:relwork-quote-extraction}), this is a ``bootstrapping'' model built from high-precision patterns, as follows.

The usage of \emph{reporting verbs} is a typical element of quote extraction models~\cite{DBLP:conf/emnlp/ParetiOKCK13}.
Along with common verbs that are used to quote others (e.g., ``say,'' ``claim'') we used verbs that are common in scientific contexts, such as ``prove'' or ``analyze.''
First, we manually create a seed set of such verbs.
Next, we extend it with their nearest neighbors in a word embedding space; the word embeddings we use are the \emph{GloVe} embeddings, which are trained on a corpus of Wikipedia articles~\cite{DBLP:conf/emnlp/PenningtonSM14}.
We follow a similar approach for nouns related to studies (e.g., ``survey,'' ``analysis'') and nouns related to scientists (e.g., ``researcher,'' ``analyst'').
Syntactically, quotes are usually expressed using indirect speech.
Thus, we also obtain part-of-speech tags from the candidate quote-containing sentences.

Using this information, we construct a series of regular expressions over \emph{classes} of words (``reporting verbs,'' ``study-related noun,'' and part-of-speech tags) which we evaluate in \S\ref{subsec:results-indicator-level}.

\spara{Quote Attribution.}
For the purposes of evaluating article quality, it is fundamental to know not only that an article has quotes, but also their provenance: \emph{who} or \emph{what} is being quoted. % (i.e., the quotee). %, i.e., the identity of the quotee.
After extracting all the candidate quote-containing sentences, we categorize them according to the information available about their quotee.

A quotee can be an \emph{unnamed scientist} or an \emph{unnamed study} if the person or article being quoted is not disclosed (e.g.,  ``researchers believe,'' ``most scientists think'' and other so-called ``weasel'' words).
%
%\footnote{These words are also known as weasel words (\url{https://en.wikipedia.org/wiki/Weasel_word}).} contained in the lists of nouns related to scientists and studies that we previously described.
%
Sources that are not specifically attributed such as these ones are as a general rule considered less credible than sources in which the quotee is named~\cite{DBLP:conf/www/ZhangRMASGAVLRB18}.

A quotee can also be a \emph{named entity} identifying a specific person or organization.
In this case, we apply several heuristics for quote attribution.
If the quotee is a \emph{named person}, if she/he is referred with her/his last or first name, we search within the article for the full name.
When the full name is not present in the article, we map the partial name to the most common full name that contains it within our corpus of news articles.
% (for instance, the last name ``Obama'' is mapped to the name of former US President ``Barrack Obama'' and not to, e.g., former US First Lady ``Michele Obama'' because the first appears more frequently in articles than the latter).
%
We also locate sentences within the article that mention this person together with a named organization. This search is performed from the beginning of the article as we assume they follow an \emph{inverted pyramid} style. In case there are several, the most co-mentioned organization is considered as the affiliation of the quotee.

If the quotee is an \emph{organization}, then it can be either mentioned in full or using an acronym.
We map acronyms to full names of organizations when possible (e.g., we map ``WHO'' to ``World Health Organization'').
If the full name is not present in an article, we follow a similar procedure as the one used to determine the affiliation of a researcher, scanning all the articles for co-mentions of the acronym and a named organization.

An illustrative example of the extraction and the attribution phase can be shown in Figure~\ref{fig:quote_example}.

\spara{Scientific Mentions.}
News articles tend to follow journalistic conventions rather than scientific ones~\cite{dunwoody2014science}; regarding citation practices, this implies they seldom include formal references in the manner in which one would find them in a scientific paper.
Often there is no explicit link: journalists may consider that the primary source is too complex or inaccessible to readers to be of any value, or may find that the scientific paper is located in a ``pay-walled'' or otherwise inaccessible repository.
However, even when there is no explicit link to the paper(s) on which an article is based, good journalistic practices still require to identify the information source (institution, laboratory, or researcher).

Mentions of academic sources are partially obtained during the quote extraction process (\S\ref{subsubsec:methods-content-quotes}), and complemented with a second pass that specifically looks for them.
During the second pass, we use the list of universities and scientific portals that we used during the \textit{crawling phase} of the data collection (\S\ref{subsec:corpus-crawling}).

\subsection{Scientific Literature Indicators}
\label{subsec:methods-papers}

In this section, we describe content- and graph-based indicators measuring how articles are related to the scientific literature.

\subsubsection{Source Adherence Indicators}
\label{subsubsec:methods-source-adherence}

\begin{figure}[t]
	\centering
	\includegraphics[width=\columnwidth]{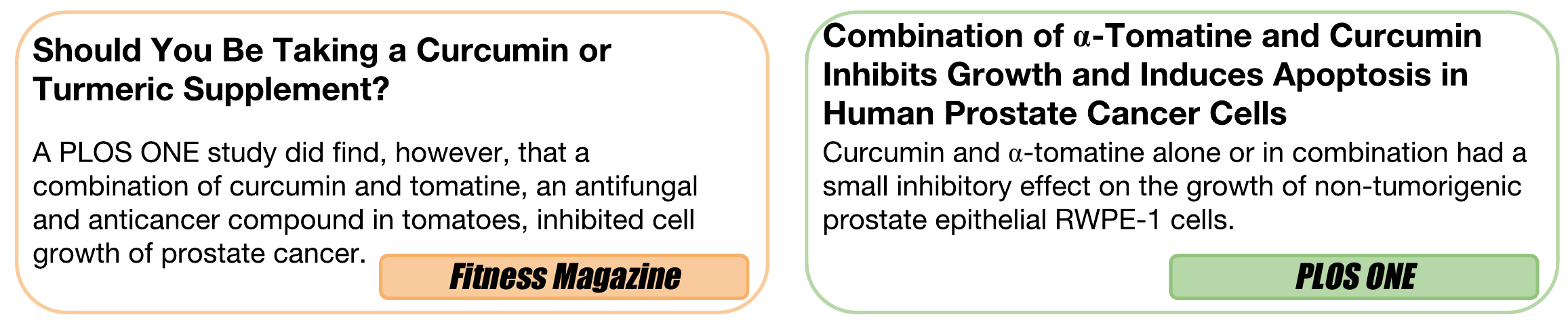}
	\caption{A news article (left) and a scientific paper (right) with Semantic Text Similarity of 87.9\%.
	%https://journals.plos.org/plosone/article?id=10.1371/journal.pone.0144293
	%https://www.fitnessmagazine.com/mind-body/supplements/should-you-be-taking-a-curcumin-or-turmeric-supplement
	%
	Indicatively, two passages from these documents, whose conceptual similarity is captured by our method, are presented.
	In these two passages we can see the effort of the journalist on translating from an academic to a less formal language, without misrepresenting the results from the paper.}
	\label{fig:similarity_example}
\end{figure}

When there is an explicit link from a news article to the URL where a scientific paper is hosted, we can measure the extent to which these two documents convey the same information.
This is essentially a computation of the \textit{Semantic Text Similarity} (STS) between the news article and its source(s).

\spara{Supervised Learning for STS.}
We construct an STS model using supervised learning.
The \textbf{features} that we use as input to the model consist of the following text similarity metrics:
\begin{enumerate*}[label=(\roman*)]
	\item the Jaccard similarity between the sets of named entities (persons and organizations), dates, numbers and percentages of the two texts;
	\item the cosine similarity between the \emph{GloVe} embeddings of the two texts;
	\item the Hellinger similarity \cite{hellinger1909neue} between topic vectors of the two texts (obtained by applying LDA \cite{DBLP:journals/jmlr/BleiNJ03}); and
	\item the relative difference between the length in words of the two texts.
\end{enumerate*}
Each of them is computed three times:
\begin{enumerate*}
	\item considering the entire contents of the article and the paper;
	\item considering one paragraph at a time, and then computing the average similarity between a paragraph in one document and a paragraph in the other; and
	\item considering one sentence at a time, and then computing the average similarity between a sentence in one document and a sentence in the other.
\end{enumerate*}
In other words, in (2) and (3) we compute the average of each similarity between the Cartesian product of the passages.

The \textbf{training data} that we use is automatically created from pairs of documents consisting of a news article and a scientific paper.
Whenever a news article has exactly one link to a scientific paper, we add the article and the paper to training data in the positive class.
For the negative class, we sample random pairs of news articles and papers.
The \textbf{learning schemes} used are Support Vector Machine, Random Forests and Neural Networks. Details regarding the evaluation of these schemes are provided in \S\ref{subsubsec:evaluation-sts}.
An example of a highly related pair of documents, as determined by this method, is shown in Figure~\ref{fig:similarity_example}.

\spara{Handling Multi-Sourced Articles.}
When an article has a single link to a scientific paper, we use the STS of them as an indicator of quality.
When an article has multiple links to scientific papers, we select the one that has the maximum score according to the STS model we just described.
We remark that this is just an indicator of article quality and we do not expect that by itself it is enough to appraise the quality of the article.
Deviations from the content of the scientific paper are not always wrong, and indeed a good journalist might consult multiple sources and summarize them in a way that re-phrases content from the papers used as sources.

\subsubsection{Diffusion Graph Indicators}
\label{subsubsec:methods-papers-graph}

We also consider that referencing scientific sources, or referencing pages that reference scientific sources, are good indicators of quality.
Figure~\ref{fig:scilens_architecture_crawling} showing a graph from scientific papers to articles, and from articles to social media postings and from them to their reactions, suggests this can be done using graph-based indicators. We consider the following:

\begin{enumerate}[nosep,leftmargin=1em,labelwidth=*,align=left]
	\item personalized PageRank~\cite{DBLP:journals/tkde/Haveliwala03} on the graph having scientific articles and universities as root nodes and news articles as leaf nodes;
	and
	\item betweenness and degree on the full diffusion graph~\cite{freeman1977set,freeman1978centrality}.
\end{enumerate}

Additionally, we consider the traffic score computed by \rurl{Alexa.com} %(\url{https://alexa.com/siteinfo})
for the website in which each article is hosted, which estimates the total number of visitors to a website.

\subsection{Social Media Indicators}
\label{subsec:methods-social}

We extract signals describing the quantity and characteristics of social media postings referencing each article.
Quantifying the amount of reactions in various ways might give us signals about the interest in different articles (\S\ref{subsec:methods-social-reach}).
However, this might be insufficient or even misleading, if we consider that false news may reach a larger audience and propagate faster than actual news~\cite{doi:10.1126/science.aap9559}.
Hence, we also need to analyze the content of these postings (\S\ref{subsec:methods-social-stance}).

\subsubsection{Social Media Reach}
\label{subsec:methods-social-reach}

Not every social media user posting the URL of a scientific news article agrees with the article's content, and not all users have sufficient expertise to properly appraise its contents. Indeed, sharing articles and reading articles are often driven by different mechanisms~\cite{DBLP:conf/cikm/AgarwalCW12}.
However, and similarly to citation analysis and to link-based ranking, the volume of social media reactions to an article might be a signal of its quality, although the same caveats apply.

Given that we do not have access to the number of times a social media posting is shown to users, we extract several proxies of the \emph{reach} of such postings.
First, we consider the total number of postings including a URL and the number of times those postings are ``liked'' in their platform.
Second, we consider the number of followers and followees of posting users in the social graph.
Third, we consider a proxy for international news coverage, which we operationalize as the number of different countries (declared by users themselves) from which users posted about an article.

Additionally, we assume that a level of attention that is sustained can be translated to a larger exposure and may indicate long-standing interest on a topic.
Hence, we consider the temporal coverage i.e., the length of the time window during which postings in social media are observed.
To exclude outliers, we compute this period for 90\% of the postings, i.e., the article's ``shelf life''~\cite{DBLP:conf/cscw/CastilloEPS14}.

\subsubsection{Social Media Stance}
\label{subsec:methods-social-stance}

\begin{figure}[t]
	\centering\small
	\includegraphics[width=\columnwidth]{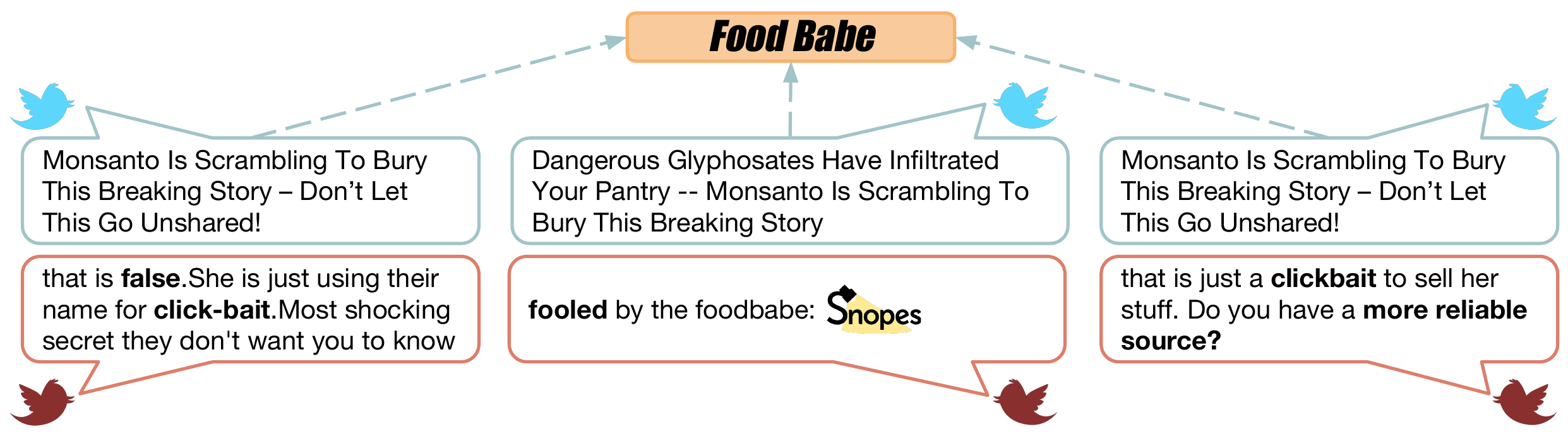}
	\caption{Example in which the stance of social media replies (bottom row) indicates the poor quality of an article promoted through a series of postings (top row).}
	\label{fig:stance_example}
\end{figure}

We consider the stance or positioning of social media postings with respect to the article they link to, as well as the stance of the responses (replies) to those postings.
According to what we observe in this corpus, repliers sometimes ask for (additional) sources, express doubts about the quality of an article, and in some cases post links to fact-checking portals that contradict the claims of the article.
These repliers are, indeed, acting as ``social media fact-checkers,'' as the example in Figure~\ref{fig:stance_example} shows.
Following a classification used for analyzing ideological debates~\cite{DBLP:conf/ijcnlp/HasanN13}, we consider four possible stances: supporting, commenting, contradicting, and questioning.

\spara{Retrieving replies.}
Twitter's API does not provide a programmatic method to retrieve all the replies to a tweet.
Thus, we use a web scraper that retrieves the text of the replies of a tweet from the page in which each tweet is shown on the web.
The design of this web scraper is straightforward and allows us to retrieve all the \textit{first-level} replies of a tweet.
%
%Thus, we cannot reconstruct the tree-like discussions with arbitrary depth regarding a news article, but only exploit their first level.

\spara{Classifying replies.}
To train our stance classifier, we use:
\begin{enumerate*}[label=(\roman*)]
\item a general purpose dataset provided in the context of \textit{SemEval 2016} \cite{DBLP:journals/toit/MohammadSK17}, and
\item a set of $300$ tweets from our corpus which were annotated by crowdsourcing workers.
\end{enumerate*}
From the first dataset we discard tweets that are not relevant to our corpus (e.g., debates on \textit{Atheism}), thus we keep only debates on \textit{Abortion} and \textit{Climate Change}.
The second set of annotated tweets is divided into $97$ contradicting, $42$ questioning, $80$ commenting and $71$ supporting tweets.
We also have $10$ tweets that were marked as ``not-related'' by the annotators and thus we exclude them from our training process.
The combined dataset contains 1,140 annotated tweets.
The \textbf{learning scheme} we use is a Random Forest classifier based on \textbf{features} including the number of:
\begin{enumerate*}[label=(\roman*)]
	\item total words,
	\item positive/negative words (using the Opinion Lexicon~\cite{DBLP:conf/kdd/HuL04}),
	\item negation words,
	\item URLs,
	and
	\item question/exclamation marks.
\end{enumerate*}
We also computed the similarity between the replies and the tweet being replied to (using cosine similarity on \emph{GloVe} vectors~\cite{DBLP:conf/emnlp/PenningtonSM14}), and the sentiment of the reply and the original tweet~\cite{StevenLoria2018}.
%
%The result is a classifier having more than 80\% accuracy when considering two classes of stances: supporting and commenting as a ``positive'' class with 151 instances; and contradicting and questioning as a ``negative class'' with 139 instances.
%
%During the training, the most popular class was sub-sampled randomly to re-balance classes.
%
Details regarding the evaluation are provided in \S~\ref{subsubsec:evaluation-stance}.

\begin{table*}[ht]
	\centering
	\caption{Summary of all the quality indicators provided by the framework SciLens.}
	\label{table:indicators_full}
	\begin{tabular}{lll}
		\toprule
		Context & Type & Indicator \\
		\midrule
		\multirow{2}{*}{Article} & Baseline & Title [Clickbait, Subjectivity, Polarity], Article Readability, Article Word Count, Article Bylined \\
		& 	Quote-Based & \#Total Quotes, \#Person Quotes, \#Scientific Mentions, \#Weasel Quotes \\
		\multirow{2}{*}{Sci. literature} & Source Adherence & Semantic Textual Similarity \\
		& Diffusion Graph & Personalized PageRank, Betweenness, [In, Out] Degree, Alexa Rank \\
		\multirow{2}{*}{Social media} & Reach & \#Likes, \#Retweets, \#Replies, \#Followers, \#Followees, [International News, Temporal] Coverage \\
		& Stance & Tweets/Replies [Stance, Subjectivity, Polarity] \\
		\bottomrule
	\end{tabular}
\end{table*}

% !TEX root = main.tex

\section{Experimental Evaluation}
\label{sec:results}

We begin the experimental evaluation by studying the performance of the methods we have described to extract quality indicators (\S\ref{subsec:results-indicator-level}).
Then, we evaluate if these indicators correlate with scientific news quality.
% (\S\ref{subsec:results-publication-level}-\S\ref{subsec:results-article-level}).
%
%We expect that sensible scientific quality indicators should be well-correlated with the quality of scientific reporting both at the article level and at the publication level.
%
First, we determine if publications that have a good (bad) reputation or track record of rigor in scientific news reporting have higher (lower) scores according to our indicators (\S\ref{subsec:results-publication-level}).
Second, we use labels from experts (\S\ref{subsec:results-article-level-expertsonly}) to compare quality evaluations done by non-experts with and without access to our indicators (\S\ref{subsec:results-article-level}).

\subsection{Evaluation of Indicator Extraction Methods}
\label{subsec:results-indicator-level}

\subsubsection{Quote Extraction and Attribution}
The evaluation of our quote extraction and attribution method (\S\ref{subsubsec:methods-content-quotes}) is based on a manually-annotated sample of articles from our corpus.
A native English speaker performed an annotation finding $104$ quotes ($37$ quotes attributed to persons, $33$ scientific mentions and $34$ ``weasel'' or unattributed quotes) in a random sample of $20$ articles.

We compare three algorithms:
\begin{enumerate*}[label=(\roman*)]
	\item a baseline approach based on regular expressions searching for content enclosed in quote marks, which is usually the baseline for this type of task;
	\item our quote extraction method without the quote attribution phase,
	and
	\item the quote extraction and attribution method, where we consider a quote as correctly extracted if there is no ambiguity regarding the quotee (e.g., if the quotee is fully identified in the article but the attribution finds only the last name, we count it as incorrect).
\end{enumerate*}

As we observed, although the baseline approach has the optimal precision, it is unable to deal with cases where quotes are not within quote marks, which are the majority ($\textbf{100\%}$ precision, $\textbf{8.3\%}$ recall).
Thus, our approach, without the quote attribution phase, improves significantly in terms of recall ($\textbf{81.8\%}$ precision, $\textbf{45.0\%}$ recall).
Remarkably, the heuristics we use for quote attribution work well in practice and serve to increase both precision and recall ($\textbf{90.9\%}$ precision, $\textbf{50.0\%}$ recall).
The resulting performance is comparable to state-of-the-art approaches in other domains (e.g., \citet{DBLP:conf/icwsm/PavlloP018} obtain $\textbf{90\%}$ precision, $\textbf{40\%}$ recall).

\subsubsection{Source Adherence}
\label{subsubsec:evaluation-sts}

We use the supervised learning method described on \S\ref{subsubsec:methods-source-adherence} to measure Semantic Text Similarity (STS).
We test three different learning models: Support Vector Machine, Random Forests and Neural Networks.
The three classifiers use similarities computed at the document, sentence, and paragraph level, and combining all features from the three levels.
Overall, the best accuracy ($\textbf{93.5\%}$) was achieved by using a Random Forests classifier and all the features from the three levels of granularity, combined.

\subsubsection{Social Media Stance}
\label{subsubsec:evaluation-stance}

We evaluate the stance classifier described in \S\ref{subsec:methods-social-stance} by performing 5-fold cross validation over our dataset.
When we consider all four possible categories for the stance (supporting, commenting, contradicting and questioning), the accuracy of the classifier is $\textbf{59.42\%}$.
This is mainly due to confusion between postings expressing a mild support for the article and postings just commenting on the article, which also tend to elicit disagreement between annotators.
%
%That happens mainly because
%\begin{inparaenum}[(i)]
%	\item we have a small annotated set of replies,
%	\item we use only textual features extracted from these replies,
%	\item it is generally difficult, even for people, to distinguish between mild supporting and commenting replies.
%\end{inparaenum}
%
Hence, we merge these categories into a ``supporting or commenting'' category comprising postings that do not express doubts about an article.
Similarly, we consider ``contradicting or questioning'' as a category of postings expressing doubts about an article; previous work has observed that indeed false information in social media tends to be questioned more often (e.g.,~\cite{DBLP:journals/intr/CastilloMP13}).
%
%Thus, in the context of our problem, these two categories can be interpreted as ``negative towards the content of the shared article'' while the other two (supporting and commenting) as ``neutral or positive''.
%
The problem is then reduced to binary classification. %, as we explain in \S~\ref{subsec:methods-social-stance},
%Τhe accuracy of this classifier on balanced classes is $\textbf{80.56\%}$.

To aggregate the stance of different postings that may refer to the same article, we compute their weighed average stance considering supporting or commenting as $+1$ (a ``positive'' stance) and contradicting or questioning as $-1$ (a ``negative'' stance).
As weights we consider the popularity indicators of the postings (i.e., the number of likes and retweets).
This is essentially a text quantification task~\cite{DBLP:journals/snam/GaoS16}, and the usage of a classification approach for a quantification task is justified because our classifier has nearly identical pairs of true positive and true  negative rates ($\textbf{80.65\%}$ and $\textbf{80.49\%}$ respectively), and false positive and false negative rates ($\textbf{19.51\%}$ and $\textbf{19.35\%}$ respectively).

\subsection{Correlation of Indicators among Portals of Diverse Reputability}
\label{subsec:results-publication-level}

We use two lists that classify news portals into different categories by reputability.
The first list, by the American Council on Science and Health~\cite{acsh2017} comprises 50 websites sorted along two axes: whether they produce evidence-based or ideologically-based reporting, and whether their science content is compelling.
The second list, by Climate Feedback~\cite{cf2017}, comprises 20 websites hosting 25 highly-shared stories on climate change, categorized into five groups by scientific credibility, from very high to very low.

\begin{figure}[ht]
	\centering
	\subfigure[]{\includegraphics[height=3.9cm]{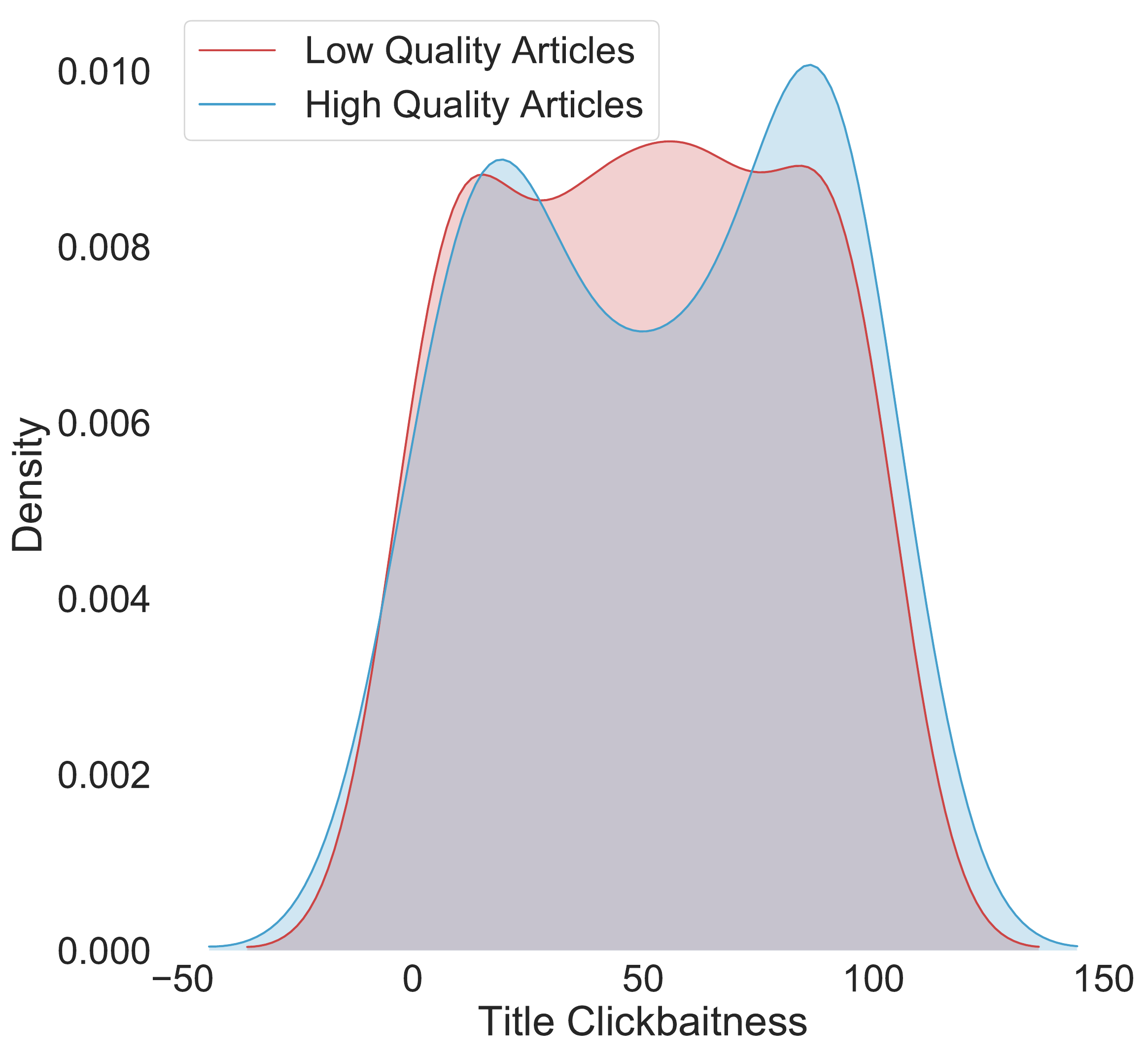}}
	\subfigure[]{\includegraphics[height=3.9cm]{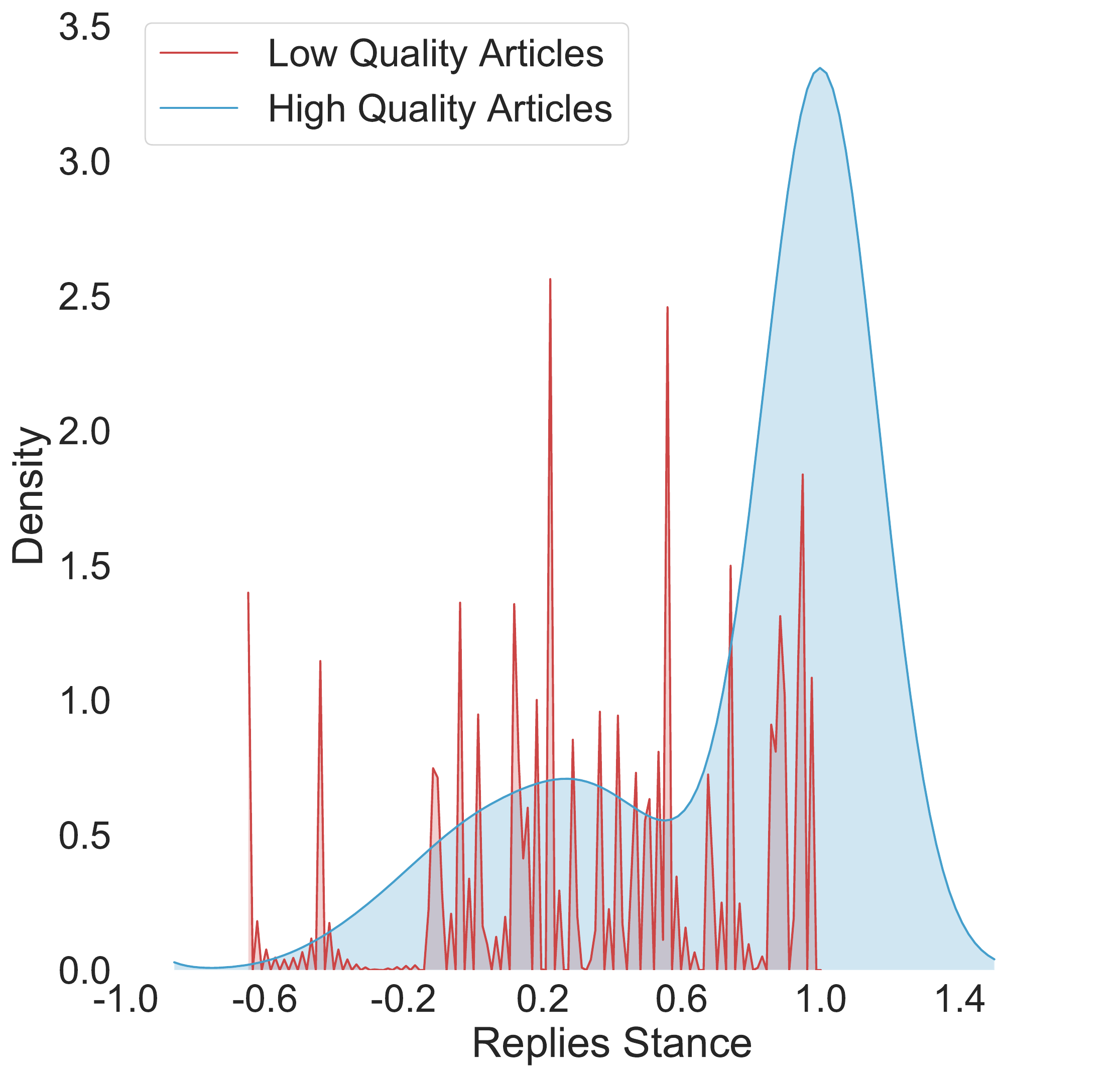}}
	\caption{Kernel Density Estimation (KDE) of a traditional quality indicator (\textit{Title Clickbaitness} on the left) and our proposal quality indicator (\textit{Replies Stance} on the right). We observe that for both high and low quality articles the distribution of \textit{Title Clickbaitness} is similar, thus the indicator is non-informative.
	However, most of the high quality articles have \textit{Replies Stance} close to $1.0$ which represents the \textit{Supporting/Commenting} class of replies, whereas low quality articles span a wider spectrum of values and often have smaller or negative values representing the \textit{Contradicting/Questioning} class of replies.
	Best seen in color.}
	\label{fig:clickbait-stance-distro}
\end{figure}

We sample a few sources according to reputability scores among the sources given consistent scores in both lists: high reputability (The Atlantic), medium reputability (New York Times), and low reputability (The Daily Mail).
Next, we compare all of our indicators in the sets of articles in our collection belonging to these sources. Two example features are compared in Figure~\ref{fig:clickbait-stance-distro}. We perform ANOVA \cite{fisher2006statistical} tests to select discriminating features. The results are shown on Table~\ref{table:indicators-importance}.
Traffic rankings by \rurl{Alexa.com}, scientific mentions, and quotes, are among some of the most discriminating features.

\begin{table}[hb]
	\centering
	\caption{Top five discriminating indicators for articles sampled from pairs of outlets having different levels of reputability (p-value: $<$ 0.005 ***, $<$ 0.01 **, $<$ 0.05 *).}
	\label{table:indicators-importance}
	\begin{tabular}{ll}
		\toprule
		The Atlantic vs. Daily Mail &    NY Times vs. Daily Mail\\
		(very high vs. very low) & (medium vs. very low) \\
		\midrule
		Alexa Rank*** & Alexa Rank***\\
		\#Scientific Mentions*** & Article Bylined***\\
		Article Readability** &  \#Scientific Mentions***\\
		\#Total Quotes* & Article Readability***\\
		Title Polarity & \#Total Quotes**\\
		\midrule\midrule
		The Atlantic vs. NY Times & All Outlets \\
		(very high vs. medium) & (from very high to very low) \\
		\midrule
		Alexa Rank*** & Alexa Rank*** \\
		Article Bylined*** & Article Bylined*** \\
		Article Word Count* & Article Word Count*** \\
		\#Replies* & \#Scientific Mentions*** \\
		\#Followers & Article Readability*** \\
		\bottomrule
	\end{tabular}
\end{table}

\subsection{Expert Evaluation}
\label{subsec:results-article-level-expertsonly}

\begin{figure*}[t]
	\centering
	\subfigure[Articles on Alcohol, Tobacco, and Caffeine]{\includegraphics[height=9cm]{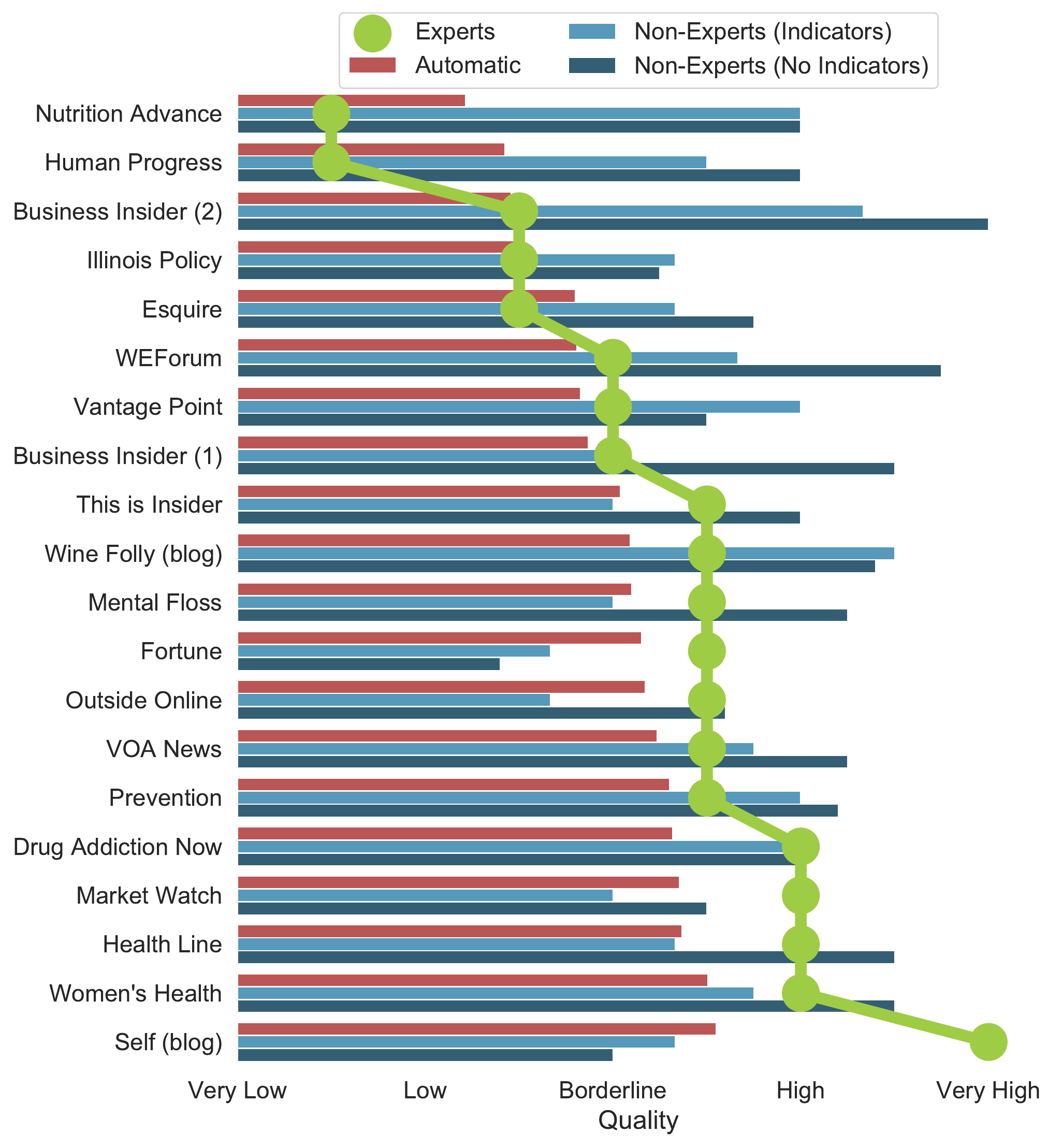}}
	\hfil
	\subfigure[Articles on CRISPR]{\includegraphics[height=9cm]{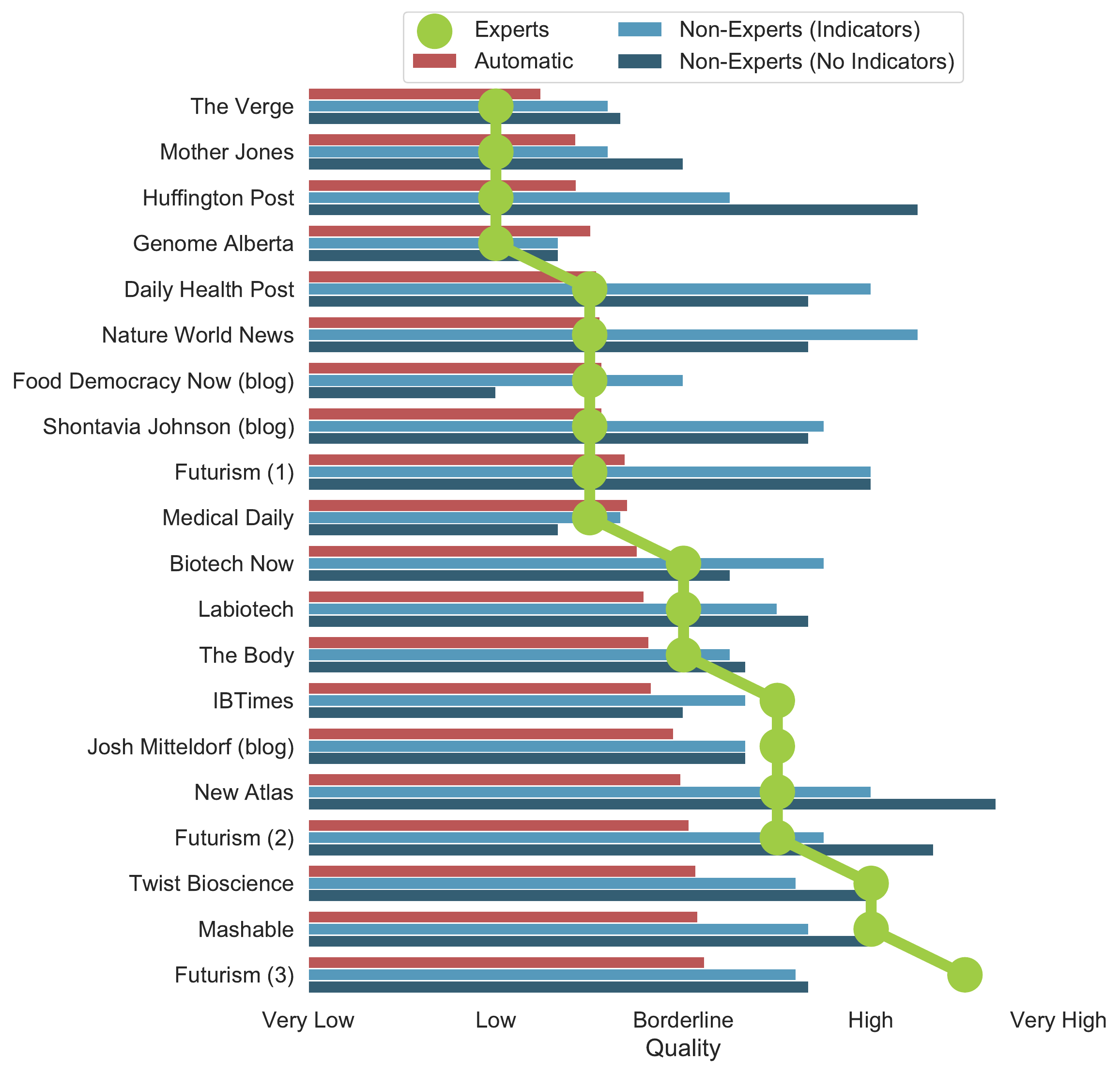}}
	\caption{Evaluation of two sets of 20 scientific articles.
	The line corresponds to expert evaluation, while the bars indicate fully automatic evaluation (red), assisted evaluation by non-experts (light blue), and manual evaluation by non-experts (dark blue). 
	Best seen in color.
	}
	\label{fig:expert-vs-non-expert-experiment}
\end{figure*}

We ask a set of four external experts to evaluate the quality of a set of articles.
%
%The experts include three people who work in communication of science in an academic context and one biologist.
%
Two of them evaluated a random sample of 20 articles about the gene editing technique CRISPR, which is a specialized topic being discussed relatively recently in mass media.
The other two experts evaluated a random sample of 20 articles on the effects of Alcohol, Tobacco, and Caffeine (the ``ATC'' set in the following), which are frequently discussed in science news.

Experts were shown a set of guidelines for article quality based on previous work (\S\ref{sec:RelatedWork}). Then, they read each article and gave it a score in a 5-point scale, from very low quality to very high quality.
Each expert annotated the 20 articles independently, and was given afterwards a chance to cross-check the ratings by the other expert and revise her/his own ratings if deemed appropriate.
%
%Detailed instructions as well as annotations are included in our data release.

The agreement between experts is distributed as follows:
\begin{enumerate*}[label=(\roman*)]
	\item \textit{strong agreement}, when the rates after cross-checking are the same (7/20 in ATC, 6/20 in CRISPR);
	\item \textit{weak agreement}, when the rates differ by one point (12/20 in ATC, 10/20 in CRISPR),
	and
	\item \textit{disagreement}, when the rates differ by two or more points (1/20 in ATC, 4/20 in CRISPR).
\end{enumerate*}
Annotation results are show on Figure~\ref{fig:expert-vs-non-expert-experiment}, and compared to non-expert evaluations, which are described next.

\subsection{Expert vs Non-Expert Evaluation}
\label{subsec:results-article-level}

We perform a comparison of quality evaluations by experts and non-experts. Non-experts are workers in a crowdsourcing platform.
We ask for five non-expert labels per article, and employ what our crowdsourcing provider, \emph{Figure Eight} (\rurl{figure-eight.com}), calls tier-3 workers, which are the most experienced and accurate. As a further quality assurance method, we use the agreement among workers to disregard annotators producing consistently annotations that are significantly different from the rest of the crowd. This is done at the worker level, and as a result we remove on average about one outlier judgment per article.

We consider two experimental conditions.
On the first condition (\textbf{non-expert without indicators}), non-experts are shown the exact same evaluation interface as experts.
On the second condition (\textbf{non-expert with indicators}), non-experts are shown $7$ of the quality indicators we produced, which are selected according to Table~\ref{table:indicators-importance}.
Each indicator (except the last two) is shown with stars, with
{\footnotesize \FiveStar\FiveStarOpen\FiveStarOpen\FiveStarOpen\FiveStarOpen} indicating that the article is in the lowest quintile according to that metric, and
{\footnotesize \FiveStar\FiveStar\FiveStar\FiveStar\FiveStar} indicating the article is in the highest quintile.
The following legend is provided to non-experts to interpret the indicators:

\smallskip
\fbox{\parbox{0.9\columnwidth}{
\scriptsize
\begin{tabular}{l}
\textbf{Visitors per day of this news website} (more visitors = more stars) \\
\textbf{Mentions of universities and scientific portals} (more mentions = more stars) \\
\textbf{Length of the article} (longer article = more stars) \\
\textbf{Number of quotes in the article} (more quotes = more stars) \\
\textbf{Number of replies to tweets about this article)} (more replies = more stars) \\
\textbf{Article signed by its author} (\Checkmark = signed,\XSolidBrush = not signed) \\
\textbf{Sentiment of the article's title} (\smiley\smiley = most positive, \frownie\frownie = most negative) \\
\end{tabular}
}}
\smallskip

Results of comparing the evaluation of experts and non-experts in the two conditions we have described are summarized in Figure~\ref{fig:expert-vs-non-expert-experiment}.
In the figure, the 20 articles in each set are sorted by increasing expert rating; assessments by non-experts differ from expert ratings, but this difference tends to be reduced when non-experts have access to quality indicators.

\begin{table}[b]
	\centering
	\caption{Differences among expert evaluations, evaluations provided by non-experts and fully automatic evaluations provided by the SciLens framework, measured using RMSE (lower is better).
ATC and CRISPR are two sets of 20 articles each. Strong agreement indicates cases where experts fully agree, weak agreement when they differed by one point, and disagreement when they differed by two or more points.
No-Ind. is the first experimental condition for non-experts, in which no indicators are shown.
Ind. is the second experimental condition, in which indicators are shown.}
	\label{table:RMSE}
	\begin{tabular}{llcccc}
		\toprule
		& \multicolumn{1}{c}{\textbf{Experts}}   & &  \multicolumn{2}{c}{\textbf{Non-Experts}} & \textbf{Fully} \\
		& \multicolumn{1}{c}{by agreement} & \# & No ind. & Ind. & \textbf{automated} \\
		\midrule
		\multirow{5}{*}{\rotatebox{90}{\textbf{ATC}}} & Strong agreement & 7  & $0.80$ & $\textbf{0.45}$ & $1.41$ \\
		& Weak agreement   & 12 & $1.28$ & $1.18$ & $\textbf{0.76}$ \\
		& Disagreement     & 1  & $0.40$ & $1.30$ & $\textbf{0.00}$ \\
		\cmidrule{2-6}
		& All articles     & 20 & $1.10$ & $\textbf{1.00}$ & $\textbf{1.00}$\\
		\midrule \midrule
		\multirow{5}{*}{\rotatebox{90}{\textbf{CRISPR}}}& Strong agreement & 6  & $1.40$ & $1.17$ & $\textbf{1.00}$ \\
		& Weak agreement   & 10 & $0.86$ & $0.76$ & $\textbf{0.67}$ \\
		& Disagreement     & 4  & $\textbf{0.96}$ & $1.22$ & $1.03$ \\
		\cmidrule{2-6}
		& All articles     & 20 & $1.96$ & $0.96$ & $\textbf{0.85}$ \\
		\bottomrule
	\end{tabular}
\end{table}
In Table~\ref{table:RMSE} we show how displaying indicators leads to a decrease in these differences, meaning that non-expert evaluations become closer to the average evaluation of experts, particularly when experts agree.
In the ATC set the improvement is small, but in CRISPR it is large, bringing non-expert scores about 1 point (out of 5) closer to expert scores.

Table~\ref{table:RMSE} and Figure~\ref{fig:expert-vs-non-expert-experiment} also includes a fully automated quality evaluation, built using a weakly supervised classifier over all the features we extracted. As weak supervision, we used the lists of sites in different tiers of reputability (\S\ref{subsec:results-publication-level}) and considered that \emph{all articles} on each site had the same quality score as the reputation of the site. Then, we used this classifier to annotate the 20 articles in each of the two sets. Results show that this achieves the \textbf{lowest} error with respect to expert annotations.

% !TEX root = main.tex

\section{Conclusions}\label{sec:conclusions}

We have described a method for evaluating the quality of scientific news articles.
This method, SciLens, requires to collect news articles, papers referenced in them, and social media postings referencing them.
We have introduced new quality indicators that consider quotes in the articles, the similarity and relationship of articles with the scientific literature, and the volume and stance of social media reactions.
The approach is general and can be applied to any specialized domain where there are primary sources in technical language that are ``translated'' by journalists and bloggers into accessible language.

In the course of this work, we developed several quality indicators that can be computed automatically, and demonstrated their suitability for this task through multiple experiments.
First, we showed several of them are applicable at the site level, to distinguish among different tiers of quality with respect to scientific news.
Second, we showed that they can be used by non-experts to improve their evaluations of quality of scientific articles, bringing them more in line with expert evaluations.
Third, we showed how these indicators can be combined to produce fully automated scores using weak supervision, namely data annotated at the site level.

\spara{Limitations.}
Our methodology requires access to the content of scientific papers and social media postings. Regarding the latter, given the limitations of the data scrapers we have used only replies to postings and not replies-to-replies. We have also used a single data source for social media postings.
Furthermore, we consider a broad definition of ``news'' to build our corpus, covering mainstream media as well as other sites, including fringe publications. 
%Developing a method specifically for mainstream news sources is possible but would probably require re-training.
Finally, our methodology is currently applicable only on English corpora. 

\spara{Reproducibility.}
Our code uses the following \texttt{Python} libraries:
\begin{enumerate*}[label=]
	\item \texttt{Pandas} and \texttt{Spark} for data management,
	\item \texttt{NetworkX} for graph processing,
	\item \texttt{scikit-learn} and \texttt{PyTorch} for ML, and
	\texttt{SpaCy}, \texttt{Beautiful Soup}, \texttt{Newspaper}, \texttt{TextSTAT} and \texttt{TextBlob} for NLP.
\end{enumerate*}
All the data, code as well as the expert and crowd annotations used in this paper are available for research purposes in \textbf{\emph{\url{http://scilens.epfl.ch}}}.

\section*{Acknowledgments}
We would like to thank the external experts who helped us on the evaluation of our framework:
\begin{enumerate*}[label=]
	\item \textbf{Aina Crosas} (Communication Officer at the Barcelona Biomedical Research Park),
	\item \textbf{Andreu Prados} (Pharmacist and Registered Dietitian Nutritionist specialized in Health Communication),
	\item \textbf{Jose F. Rizo} (Biologist and President of the Fundaci\'{o}n \~{N}amku), and
	\item \textbf{Dimitra Synefiaridou} (PhD Student in Microbiology at the Veening Lab).
\end{enumerate*}
This work is partially supported by the Open Science Fund of EPFL (\url{http://www.epfl.ch/research/initiatives/open-science-fund}) and the La Caixa project (LCF/PR/PR16/11110009). 

%%%%%%%%%%%%%%%%%%%%%%%%%%%%%%%%%%%%%%%%%%%%%%%%%%%%%%%%%%
% BIBLIOGRAPHY
%%%%%%%%%%%%%%%%%%%%%%%%%%%%%%%%%%%%%%%%%%%%%%%%%%%%%%%%%%

% WWW 2019 allows unlimited pages of bibliography
\clearpage
\balance
\bibliographystyle{ACM-Reference-Format}
\bibliography{references}

%%%%%%%%%%%%%%%%%%%%%%%%%%%%%%%%%%%%%%%%%%%%%%%%%%%%%%%%%%
% APPENDICES
%%%%%%%%%%%%%%%%%%%%%%%%%%%%%%%%%%%%%%%%%%%%%%%%%%%%%%%%%%

%\input{removed}

\end{document}